\documentclass[doublecol]{epl2} 

\usepackage{graphicx}
\title{The Abelian Higgs model and a minimal length in an un-particle scenario}
\shorttitle{The Abelian Higgs model and minimal length in an un-particle scenario}

\author{Patricio Gaete\inst{1} \and Euro Spallucci\inst{2}}
\shortauthor{P. Gaete \etal}

\institute{ 
\inst{1} Departmento de F\'{i}sica and Centro Cient\'{i}fico-Tecnol\'ogico de Valpara\'{i}so, 
Universidad T\'{e}cnica Federico Santa Mar\'{i}a, Valpara\'{i}so, Chile\\
  \inst{2} Dipartimento di Fisica Teorica, Universit\`a di \
Trieste and INFN, Sezione di Trieste, Italy                
}


\abstract{We consider both the Abelian Higgs model and the impact
of a minimal length in the un-particle sector. It is shown that even
if the Higgs field takes a non-vanishing
v.e.v., gauge interaction keeps its  long range character leading to
an effective gauge symmetry restoration. The effect of a quantum
gravity induced minimal length on a physical observable is then
estimated by using a physically-based alternative to the usual
Wilson loop approach. Interestingly, we obtain an ultraviolet finite
interaction energy described by a confluent hyper-geometric function, 
which shows a remarkable richness of behavior.}

\begin{document}

\maketitle

\section{Introduction}

The subject of un-particle physics has been of great interest
since the work by Giorgi \cite{Georgi:2007ek,Georgi:2007si},
who showed the possibility of describing unusual properties
of matter with non-trivial scale invariance in the infrared regime. 
The physical consequences of this new physics have triggered
a large body of literature. For example,
in connection to supersymmetry \cite{Nakayama:2007qu,Zhang:2007ih},
spin-dependent in electron systems \cite{Liao:2007ic}, also in dark
matter scenarios \cite{Kikuchi:2007az,Deshpande:2007jy}, in black
holes studies
\cite{Mureika:2007nc,Mureika:2008dx,Mureika:2009ii,Mureika:2010je,Gaete:2010sp},
holography \cite{Ho:2009zv}, and possible signatures of un-particles
interactions in collider physics
\cite{Mohanta:2007uu,Alan:2007ss,Aliev:2009sd,Kumar:2007af}. 
More recently, the un-particle phenomenology has gained a renewed
attention for the concrete possibility of setting the un-particle scale in
an unambiguous way, i.e., independent of the coupling constant between
the BZ fields and the Standard Model \cite{Frassino}. The
advent of un-particles physics has also drawn attention to the Higgs
phenomenology 
\cite{Fox:2007sy,Deshpande:2007jy,Delgado:2007dx,Kikuchi:2007qd}, 
or rather to the dynamical breaking of electroweak symmetry and its
relation with the un-particle sector \cite{Sannino:2009za}.
Here it is important to emphasize a distinctive feature of this new sector,
that is, the existence of long-range forces between particles mediated by
un-particles, which are valid in three, two and one space dimensions. 

On the other hand, the investigation of extensions of the Standard
Model (SM), such as Lorentz invariance violation and fundamental
length, have also been intensively considered by many authors
\cite{AmelinoCamelia:2002wr,Jacobson:2002hd,Konopka:2002tt,Hossenfelder:2006cw,Nicolini:2008aj,Euro}. Evidently, all this activity is related to the fact
that the SM does not include a quantum theory of gravitation. 
As is well-known, in the search for a more
fundamental theory going beyond the SM string theories \cite{Witten}
are the only known candidate for a consistent, ultraviolet finite
quantum theory of gravity, unifying all fundamental interactions.
We further note that, during the last years, the focus of quantum gravity has
been towards effective models. Indeed, these models incorporate one
of the most important and general features, that is, a \emph{minimal
length scale} that acts as a regulator in the ultraviolet. 
There are various approaches on how to construct a quantum
field theory that incorporates a minimal length scale, which gives
rise to a model of quantum space-time
\cite{Magueijo:2001cr,Hossenfelder:2003jz,Nicolini:2010bj,Sprenger:2012uc}.
Of these, quantum field theory allowing non-commuting position
operators have enjoyed great popularity
\cite{Witten:1985cc,Seiberg:1999vs,Douglas:2001ba,Szabo:2001kg,Gomis:2000sp,Bichl:2001nf}. Notice that most of the known results in the
non-commutative approach have been achieved using a
Moyal star-product. More recently, a new formulation
of  non-commutative quantum field theory in the presence of a
minimal length has been proposed in
\cite{Smailagic:2003rp,Smailagic:2003yb,Smailagic:2004yy}.
Subsequently, this was developed by the introduction of a new
multiplication rule, which is known as Voros star-product. 
Evidently, physics turns out be independent from the choice of the
type of product \cite{Hammou:2001cc}. Accordingly, with the
introduction of non-commutativity by means of a minimal length, the
theory becomes ultraviolet finite. In this case, the cutoff is
provided by the non-commutative parameter, $\theta$, which is
related to the minimal length. It should, however, be noted here that
this cut-off is not put by hand, as a mere artifact of regularization, 
but is a direct consequence of the coherent states approach 
developed in \cite{Smailagic:2003rp,Smailagic:2003yb,Smailagic:2004yy}.

In this perspective, the purpose of this paper is to
further elaborate on the physical content of un-particle physics.
Specifically, in this work we will focus attention on the impact of a Higgs
field with non-vanishing v.e.v. and a minimal length on a physical
observable. Although a preliminary analysis about
these issues has appeared before \cite{Gaete:2012cs}, we think it is of value
to clarify them because, in our view, they have not been properly
emphasized. In this way, we will work out the static
potential between two charges for the cases under consideration. As
we shall see, in the case of a Higgs field with non-vanishing v.e.v,
gauge interaction keeps its long range character leading to an
effective gauge symmetry restoration. While in the presence of a
minimal length, our analysis leads to an ultraviolet finite static
potential for non-commutative un-particles induced by a confluent
hypergeometric function, which shows a remarkable richness of behavior. 

\section{The Abelian Higgs model in the un-particle sector}

As already expressed, we now examine the properties of the Higgs
vacuum in the un-particle sector. For this purpose we restrict our attention
to scalar electrodynamics:
\begin{equation}
Z\left[\, j\,\right]=\int D\left[\, \phi\,\right]\, D\left[\,
\phi^\ast\, \right]\, D\left[\, A\,\right]\, \exp\left(\, -S\left[\,
\phi\ ,\phi^\ast\ , A\ ; j\,\right]\,\right),
\end{equation}
where the Lagrangian density is given by
\begin{equation}
{\mathcal L}= -\frac{1}{4} \, F_{\mu\nu}\, F^{\mu\nu} -\vert\,
\left(\,
\partial_\mu - i\, e\, A_\mu \,\right)\,\phi\,\vert^2 - V_H\left(\,
\phi^\ast \phi\,\right)  -j^\mu A_\mu.
 \end{equation}
$j^\mu $ is a divergence-free \emph{external} current supported by
heavy test-particles. Its role will become clear in a while. We do
not choose any specific form of the Higgs potential $V_H$, and
simply assume it admits a non-trivial minimum for a non-vanishing
v.e.v. of the field. Next, by decomposing  the $\phi$ field in a
polar form
\begin{equation}
\phi\left(\, x\,\right)=\frac{\rho\left(\, x\,\right)}{\sqrt 2}\,
e^{ i\vartheta\left(\, x\,\right)},
\end{equation}
we obtain the Higgs Lagrangian,   in terms of the modulus $\rho$ and
the phase $\vartheta$,
\begin{equation}
{\mathcal L}_H \equiv -\frac{1}{2}\left[\, \left(\, \partial_\mu -
i\, e\, A_\mu + i \,\partial_\mu\, \vartheta \,\right)\,\rho \,
\right]^2 - V_H\left(\, \rho^2\,\right),
\end{equation}
The ground state corresponds to a constant value of the $\rho$
field, say $\sigma_0$,  which minimizes the potential $V_H$
\begin{equation}
 \left(\, \frac{\partial V_H}{\partial \rho}\,\right)_{\rho=\sigma_0}=0.
\end{equation}
We define  the fluctuation field $\eta$ as $\rho\left(\,
x\,\right)\equiv \sigma_0 + \eta\left(\, x\,\right)$.  By neglecting
higher order contributions, we get to the zero-th order in $\eta$
\begin{equation}
{\mathcal L}\approx  -\frac{1}{4} \, F_{\mu\nu}\, F^{\mu\nu}
-\frac{e^2 \sigma_0^2}{2}\, \left(\,  A_\mu
-\frac{1}{e}\partial_\mu\, \vartheta\,\right)^2   -j^\mu\, A_\mu,
\end{equation}
leading to a system of  effective field equations for a massive
vector field and a Goldstone boson
\begin{eqnarray}
&&  \partial_\mu \, F^{\mu\nu} - e^2 \sigma_0^2 \left(\,  A_\mu
-\frac{1}{e}
\partial_\mu\, \vartheta\,\right)=j^\nu\ ,
\label{proca}\\
&& \partial^2\, \vartheta= e\, \partial_\mu \, A^\mu\ \
.\label{goldstone}
\end{eqnarray}
The ``current wisdom'' would suggest to switch to the unitary gauge,
where the Goldstone boson is ``eaten-up'' by the gauge field which
turns into a massive objects. We depart from this approach for two
reasons:

{\bf i}) it is not necessary to choose any gauge to proceed;

{\bf ii}) appealing to the unitary gauge gives the feeling that the whole
Higgs mechanism is a skillful ``illusion game'', not a physical
mechanism. In other words, the mass of the vector boson is a
measurable quantities which cannot depend on a clever gauge choice.

Rather than choosing a gauge, we \emph{solve} equation
(\ref{goldstone}) and write $\vartheta$ in terms of $A$:
\begin{equation}
\vartheta =e\, \frac{1}{\partial^2}\,\partial_\mu\ A^\mu.
\label{goldstone2}
\end{equation}
In equation (\ref{goldstone2}) we use the following short-hand
notation
\begin{equation}
\frac{1}{\partial^2}\, F \equiv \int d^4y\, \frac{1}{\partial^2_x}
\,\delta\left(\, x-y\,\right)\, F\left(\, y\,\right).
\end{equation}
We notice that solving for $\vartheta $ makes $A_\mu$
divergence-free:
\begin{equation}
A_\mu -\frac{1}{e}\partial_\mu\, \vartheta = \left(\,
\delta^\lambda_\mu - \frac{\partial^\lambda\,
\partial_\mu}{\partial^2} \,\right)\, A_\lambda \equiv A_\mu^\perp.
\end{equation}
Moreover,
\begin{equation}
\left(\, A_\mu^\perp\,\right)^2=\frac{1}{4}\,
F_{\mu\nu}\,\frac{1}{\partial^2}\,F^{\mu\nu}.
\end{equation}
Thus, by discarding a total divergence we get the \emph{effective
Lagrangian} for $A_\mu$ in the Higgs vacuum
\begin{equation}
{\mathcal L}^{eff.}=  -\frac{1}{4} \, F_{\mu\nu}\,\left(\, 1  -e^2
\sigma_0^2\, \frac{1}{\partial^2}\, \right)\, F^{\mu\nu} -j^\mu
A_\mu. \label{leffv}
\end{equation}
The effective Lagrangian (\ref{leffv}):

{\bf i}) is manifestly gauge invariant, notwithstanding the presence of a
mass term;

{\bf ii}) has the same form of the Schwinger model effective Lagrangian
once the electron field is integrated out and the axial anomaly
properly accounted. From this vantage point, we can see ${\mathcal
L}^{eff.} $ as the $4D$ equivalent of the Schwinger model, i.e.
$QED_2$.

In order to turn the Higgs field into its un-particle counterpart we
replace the functional integration measure as follows:
\begin{eqnarray}
\int D\left[\, \phi\,\right]\, D\left[\,
\phi^\ast\,\right]&\longrightarrow& \frac{A_{d_U}\, e^{2d_U
-2}}{2\Lambda_U^{2d_U-2} } \nonumber\\
&\times&\int_0^\infty d\sigma_0^2 \left(\,\sigma_0^2\,\right)^{d_U- 2}
\int D\left[\, \eta\,\right], \nonumber\\
\end{eqnarray}
where, $d_U$ is the non-integral scale dimension of the Higgs
un-particle field, and
\begin{equation}
A_{d_U}\equiv \frac{16\pi^{5/2}}{\left(\, 2\pi\,\right)^{2d_U}}\,
\frac{\Gamma\left(\, d_U+ 1/2\,\right)}{\Gamma\left(\, d_U-
1\,\right)\, \Gamma\left(\, d_U\,\right)} ,
\end{equation}
is a characteristic normalization factor for un-particle objects.
$\Lambda_U$ is the energy scale where the Banks-Zaks fields turn
into un-particle fields through dimensional transmutation.

Integration over the Higgs fields vacuum mean value is suggested by
the general prescription, implemented in \cite{Gaete:2008wg}, to
recover un-particle generating functional form the standard one by
properly integrating over the mass parameter. As the vacuum
expectation value of $\phi$ determines the mass of the Higgs boson,
self-consistency requires to integrate over it in the same way
\begin{eqnarray}
Z_U\left[\, j\,\right]&=&\,\frac{A_{d_U}\, e^{2d_U
-2}}{2\Lambda_U^{2d_U-2} } \int_0^\infty d\sigma_0^2 \left(\,
\sigma_0^2\,\right)^{d_U- 2} \nonumber\\
&\times& \int D\left[\, \eta\,\right] D\left[\, A\,\right]\,
\exp\left(\, -\int d^4x\left[\, L^{eff.} +
L_{int}\,\right]\,\right) \nonumber\\
\label{unz}
\end{eqnarray}

As we are interested into the static potential between test charges,
we neglect the interaction term in (\ref{unz}) and consider only the
"free" dynamics of the gauge field encoded into $L^{eff.} $. In
order to simplify the notation, let us re-define the generating
functional as
\begin{equation}
Z_U\left[\, j\,\right]\equiv \frac{A_{d_U}\, e^{2d_U
-2}}{2\Lambda_U^{2d_U-2} } \int_0^\infty d\sigma_0^2 \left(\,
\sigma_0^2\,\right)^{d_U- 2} \,
Z_A\left[\, j\,\right]\ ,\label{zunz}
\end{equation}
\begin{equation}
Z_A\left[\, j\,\right] \equiv   Z_A^{-1}\left[\, 0\,\right]\,
\int  D\left[\, A\,\right]\, \exp\left(\, -\int d^4x \,
L^{eff.}\left[\, A\ , j\,\right] \,\right). \label{zunz2}
\end{equation}
For "heavy", opposite, test charges sitting in $\vec{x}_1$,
$\vec{x}_2$ the current is $ j^\mu=  j^0\, \delta_0^\mu $, where
\begin{equation}
j^0\left(\, \vec{x}\,\right) = e\,\delta\left[\, \vec{x}-\vec{x}_1\,
\right]-e\,\delta\left[\, \vec{x}-\vec{x}_2\,\right],
\label{current}
\end{equation}
and the static  interaction between the pair is mediated by $A_0$.
Thus,
\begin{equation}
L^{eff.}=  -\frac{1}{4} \, F_{\mu\nu}\,\left(\, 1  -e^2 \sigma_0^2\,
\frac{1}{\partial^2}\, \right)\, F^{\mu\nu} -j^0\, A_0.
\label{leffstat}
\end{equation}
As we are dealing with a static problem, the Hamiltonian formalism
is appropriate to extract the static potential from the canonical
path integral.

The canonical momentum conjugated to $ A_i $ is the \emph{electric
field}  $E^i$:
\begin{equation}
\frac{\partial L^{eff.}}{\partial \partial_0 \, A_i }= - \left(\, 1
-e^2 \sigma_0^2\,\frac{1}{\partial^2} \, \right)\, F^{0\, i} \equiv
E^i,
\end{equation}
and the Hamiltonian turns out to be
\begin{equation}
H= \frac{1}{2}\, E_i\, \left(\, 1  -e^2
\sigma_0^2\,\frac{1}{\partial^2}\, \right)^{-1}\, E^i +
E^i\,\partial_i \, A_0 +j^0\, A_0.
\end{equation}
The action reads
\begin{eqnarray}
S^{eff.}&=&\int d^4x \,\left[\, -A_i \,\partial_0 \, E^i
-\frac{1}{2}\, E_i\, \left(\, 1  -e^2
\sigma_0^2\,\frac{1}{\partial^2}\, \right)^{-1}\, E^i\,\right] \nonumber\\
&+&\int d^4x\,\left[A_0\, \left(\,\partial_i\, E^i -j^0\,\right)
\,\right]. \label{seff}
\end{eqnarray}
The form of  (\ref{seff}) shows that both  $A_i$ and $A_0$ are
Lagrange multipliers implementing  the static nature of the electric
field and the Gauss Law:
\begin{eqnarray}
&& \frac{\delta S^{eff.}}{\delta A_i}=0 \longrightarrow \partial_0 \vec{E}=0\ ,\\
&& \frac{\delta S^{eff.}}{\delta A_0}=0 \longrightarrow
\vec{\nabla}\cdot \vec{E}=j^0\ .
\end{eqnarray}
A straightforward computation of the path integral shows that
\begin{equation}
Z_A\left[\, j^0\,\right] =\exp\left( -\frac{1}{2}\int d^4x \
j^0\, \frac{1}{ \nabla^2  -e^2 \sigma_0^2}\, j^0\,\right)\,,
\end{equation}
and
\begin{equation}
V\left(   \,L\,\right)=-\frac{e^2}{4\pi\, L} \, e^{ - e\,\sigma_0
\, L }\,,
\end{equation}
where$\qquad L\equiv\vert \vec{x}_1 -\vec{x}_2\,\vert$.

Up to now we have reproduced known results but in a \emph{gauge
invariant way} at every step of calculation.

Now, let us define the static potential in the un-particle sector of
the model as
\begin{equation}
Z_U\left(\, L\,\right) = \exp\left( - T \, V_U\left(   \,L\,\right)
\,\right)\label{VU}.
\end{equation}
Comparing (\ref{VU}) and (\ref{zunz}) we see that
\begin{equation}
V_U\left(   \, L \,\right)=\frac{A_{d_U}\, e^{2d_U
-2}}{2\Lambda_U^{2d_U-2} } \int_0^\infty d\sigma_0^2 \left(\,
\sigma_0^2\,\right)^{d_U- 2} \, V\left(   \, L \,\right).
\end{equation}
By introducing the new integration variable $ t\equiv e^2
\,\sigma_0^2 $, $V_U\left(   \, L \,\right) $ can be written as
\begin{equation}
V_U\left(   \, L \,\right)=-\frac{ e^2 }{8\pi}\, A_{d_U}\, \frac{1 }
{\Lambda_U^{2d_U-2} } \int_0^\infty dt\, t^{d_U- 2} \,
e^{-L\,\sqrt{t}},
\end{equation}
The final form of $V_U$ results to be
\begin{equation}
V_U\left(   \, L \,\right)=-\frac{ e^2 }{4\pi\, L}\, \Gamma\left(\,
2d_U-2\,\right)\, \frac{A_{d_U} }{\left(\,
L\,\Lambda_U\,\right)^{2d_U-2} }.  \label{vuf}
\end{equation}
Equation (\ref{vuf}) shows that even if the Higgs field has a non
vanishing v.e.v. in the un-particle vacuum, gauge interaction is
still long-range.

\section{A minimal length in the un-particle sector}

We shall now calculate the interaction energy between test charges in the
un-particle sector introducing a quantum gravity induced universal
cut-off. For this purpose, we will compute the expectation value of the
energy operator $H$ in the physical state $|\Phi\rangle$ describing
the sources, which we will denote by $ {\langle H\rangle}_\Phi$. With
this in view, the initial point of our analysis is the Lagrangian
density:
\begin{equation}
\mathcal{L} = \sum\limits_{k = 1}^N {\left[ { - \frac{1}{{4e_k^2 }}
F^{k\mu \nu } F_{\mu \nu }^k  + \frac{{m_k^2 }}{{2e_k^2 }}\left(
{A_\mu ^k  - \partial _\mu  \varphi ^{k} } \right)^2 } \right]},
\label{NCunp05}
\end{equation}
where $m_{k}$ is the mass for the $N$ scalar fields. As we have
already indicated in \cite{Gaete:2008wg,Gaete:2008aj}, integrating
out the $\varphi$-fields induces an effective theory for the
$A_{\mu}^{k}$-fields. Accordingly, we obtain the following effective
Lagrangian density:
\begin{equation}
\mathcal{L} = \sum\limits_{k = 1}^N {\frac{1}{{e_k^2 }}} \left[ { -
\frac{1}{4}F_{\mu \nu }^k \left( {1 + \frac{{m_k^2 }}{\Delta }}
\right)F^{k\mu \nu } } \right]. \label{NCunp10}
\end{equation}
By proceeding in the same way as in
\cite{Gaete:2008wg,Gaete:2008aj}, we obtain the Hamiltonian
\begin{eqnarray}
H  &=& \int {d^3 x} \left\{ { -\frac{1}{2}\Pi ^i \left( {1 +
\frac{{m_k^2 }}{\Delta }} \right)^{ - 1} \Pi _i} \right\}\nonumber\\
&+&\int {d^3 x} \left\{ {\frac{1}{4} F_{ij}
\left( {1 + \frac{{m_k^2 }} {\Delta }} \right)F^{ij}} \right\}.
\label{NCunp20}
\end{eqnarray}
Next, following our earlier Hamiltonian procedure, we recall that
\begin{eqnarray}
\Pi _i \left( x \right)\left| {\overline \Psi  \left( {\bf y}
\right)\Psi \left( {\bf y^\prime} \right)} \right\rangle  &=&
\overline \Psi \left( {\bf y} \right)\Psi \left( {\bf y^\prime}
\right)\Pi _i \left( x \right)\left| 0 \right\rangle \nonumber\\
&+&q\int\limits_{\bf y}^{\bf y^\prime} {dz_i \delta ^{(3)} \left( {{\bf
z} - {\bf x}} \right)\left| \Phi \right\rangle }.
\label{NCunp40a}
\end{eqnarray}
As was explained in 
\cite{Gaete:2011ka,Gaete:2012yu, Smailagic, Moffat, Nicolini}, we now
consider the formulation of this theory in the presence of a minimal
length. To do this, we replace the source by the smeared one:
\begin{equation}
\delta ^{\left( 3 \right)} \left( {x - y} \right) \to
e^{{\textstyle{\theta  \over 2}} \nabla ^2 } \delta ^{\left( 3
\right)} \left( {x - y} \right). \label{NCunp40a1}
\end{equation}
Hence expression $(\ref{NCunp40a})$ reduces to
\begin{eqnarray}
\Pi _i \left( x \right)\left| {\overline \Psi  \left( {\bf y}
\right)\Psi \left( {\bf y^\prime} \right)} \right\rangle  &=&
\overline \Psi \left( {\bf y} \right)\Psi \left( {\bf y^\prime}
\right)\Pi _i \left( x \right)\left| 0 \right\rangle \nonumber\\
&+&q\int\limits_{\bf y}^{\bf y^\prime} {dz_i e^{{\textstyle{\theta
\over 2}}\nabla ^2 }     \delta ^{(3)} \left( {{\bf z} - {\bf x}}
\right)\left| \Phi \right\rangle }. \nonumber\\
\label{NCunp40b}
\end{eqnarray}
Having made this observation and since the fermions are taken to be
infinitely massive (static) we can substitute $\Delta$ by $- \nabla
^2$ in  Eq. $(\ref{NCunp20})$. Accordingly, $\left\langle H
\right\rangle _\Phi$ takes the form
\begin{equation}
\left\langle H \right\rangle _\Phi   = \left\langle H \right\rangle
_0  + V, \label{NCunp45}
\end{equation}
where $\left\langle H \right\rangle _0  = \left\langle 0
\right|H\left| 0 \right\rangle$. The $V$ term is given by:
\begin{eqnarray}
V &=& \frac{{q^2 }}{2}\int_y^{y^ \prime  } {dz_i^ \prime }
\frac{{\nabla _{z^ \prime  }^2 }}{{\left( {\nabla ^2  - m_k^2 }
\right)_{z^ \prime  } }} \nonumber\\
&\times&\int_y^{y^ \prime  } {dz^i } \left( { -
e^{\theta \nabla _z^2 } \delta ^{\left( 3 \right)} \left( {z - z^
\prime  } \right)} \right), \label{NCunp50}
\end{eqnarray}
where the integrals over $z^i$ and $z^\prime_i$ are zero except on
the contour of integration. Expression (\ref{NCunp50}) can also be
written as
\begin{equation}
V = \frac{{q^2 }}{2}\int_{\bf y}^{{\bf y}^{\prime}  }
{dz_i^{\prime}}\partial _i^{z^{\prime}} \int_{\bf y}^{{\bf
y}^{\prime}} {dz^i }\partial _z^i \tilde G\left( {{\bf
z}^{\prime},{\bf z}} \right), \label{NCunp55}
\end{equation}
where $\tilde G$ is the Green function
\begin{equation}
\tilde G\left( {{\bf z} - {\bf z}^ \prime  } \right) = \int_0^\infty
{ds} e^{ - sM^2 } \left( {\frac{1}{{4\pi \left( {\theta  + s}
\right)}}} \right)^{{\raise0.5ex\hbox{$\scriptstyle 3$}
\kern-0.1em/\kern-0.15em \lower0.25ex\hbox{$\scriptstyle 2$}}}
{\mathop{\rm e}\nolimits} ^{ - \frac{{r^2 }}{{4\left( {\theta  + s}
\right)}}} , \label{NCunp60}
\end{equation}
where $r\equiv|{\bf z}-{\bf {z^\prime}}|$. We notice that the
integrand in equation (\ref {NCunp60}) is just the flat space counterpart of
the regular kernel introduced in \cite{Smailagic}, where it was instrumental
to determine the trace anomaly in a "quantum" manifold.

Employing Eq.(\ref{NCunp60}) and remembering that the integrals over
$z^i$ and $z_i^{\prime}$ are zero except on the contour of
integration, expression (\ref{NCunp55}) reduces to a finite Yukawa
interaction. Therefore the potential for two opposite charges
located at ${\bf y}$ and ${\bf {y^\prime}}$ is given by
\begin{equation}
V=-\frac{{q^2 }}{{4\pi ^{{\raise0.5ex\hbox{$\scriptstyle 3$}
\kern-0.1em/\kern-0.15em \lower0.25ex\hbox{$\scriptstyle 2$}}}
}}\int_0^\infty  {ds} e^{ - sM^2 } \left( {\frac{1}{{4\pi \left(
{\theta  + s} \right)}}} \right)^{{\raise0.5ex\hbox{$\scriptstyle
3$} \kern-0.1em/\kern-0.15em \lower0.25ex\hbox{$\scriptstyle 2$}}}
{\mathop{\rm e}\nolimits} ^{ - \frac{{r^2 }}{{4\left( {\theta  + s}
\right)}}} , \label{NCunp65}
\end{equation}
where $L\equiv|{\bf y}-{\bf {y^\prime}}|$. However, from
(\ref{NCunp10}) we must sum over all the modes in (\ref{NCunp65}),
that is,
\begin{eqnarray}
V &=&  - \frac{{q^2 }}{{\left( {4\pi }
\right)^{{\raise0.5ex\hbox{$\scriptstyle 3$}
\kern-0.1em/\kern-0.15em \lower0.25ex\hbox{$\scriptstyle 2$}}}
}}\sum\limits_{k = 1}^N {\frac{1} {{e_k^2 }}} \nonumber\\
&\times&\int_0^\infty  {ds}
e^{ - sm_k^2 } \left( {\frac{1}{{4\pi \left( {\theta  + s}
\right)}}} \right)^{{\raise0.5ex\hbox{$\scriptstyle 3$}
\kern-0.1em/\kern-0.15em \lower0.25ex\hbox{$\scriptstyle 2$}}}
{\mathop{\rm e}\nolimits} ^{ - \frac{{r^2 }} {{4\left( {\theta + s}
\right)}}}. \label{NCunp66}
\end{eqnarray}
In effect, as was explained in Ref. \cite{Krasnikov:2007fs}, in the
limit $k \to\infty$ the sum is substituted by an integral as follows
\begin{eqnarray}
V &=&  - \frac{{q^2 }}{{\left( {4\pi }
\right)^{{\raise0.5ex\hbox{$\scriptstyle 3$}
\kern-0.1em/\kern-0.15em \lower0.25ex\hbox{$\scriptstyle 2$}}}
}}\frac{{A_{d_U } }} {{\Lambda _U^{2d_U  - 2} }}\int_0^\infty  {dt}
t^{d_U  - 2} \nonumber\\
&\times&\int_0^\infty  {ds} e^{ - st} \frac{1}{{\left( {\theta +
s} \right)^{{\raise0.5ex\hbox{$\scriptstyle 3$}
\kern-0.1em/\kern-0.15em \lower0.25ex\hbox{$\scriptstyle 2$}}}
}}{\mathop{\rm e} \nolimits} ^{ - \frac{{r^2 }}{{4\left( {\theta  +
s} \right)}}}, \label{NCunp70}
\end{eqnarray}
with $t=m_{k}^{2}$ and $\Lambda _U$ is a critical energy scale below
which the standard model particles can interact with un-particles.
Here ${t^{d_U  - 2} }$ is the spectral density, and $A_{d_{U}}$ is a
normalization factor which is given by
\begin{equation}
A_{d_U}\equiv \frac{16\,\pi^{5/2}}{\left(\, 2\pi\,\right)^{2d_U}}\,
\frac{\Gamma\left(\, d_U +1/2\,\right)}{\Gamma\left(\, d_U
-1\,\right)\,
 \Gamma\left(\, 2d_U \,\right)}. \label{NCunp75}
\end{equation}
Thus  one has for the interaction energy the result:
\begin{eqnarray}
V &=&  - \frac{{q^2 }}{{\left( {2\pi } \right)^{2d_U  - 1} }}\frac{1}
{{\Lambda ^{2d_U  - 2} }}\frac{1}{{\left( \theta \right)^{d_U  -
{\raise0.5ex\hbox{$\scriptstyle 1$} \kern-0.1em/\kern-0.15em
\lower0.25ex\hbox{$\scriptstyle 2$}}} }}\frac{{\Gamma \left( {d_U  +
{\raise0.5ex\hbox{$\scriptstyle 1$} \kern-0.1em/\kern-0.15em
\lower0.25ex\hbox{$\scriptstyle 2$}}} \right)}}{{\Gamma \left( {2d_U
} \right)}} \nonumber\\
&\times&\Phi \left( {d_U  - {\raise0.5ex\hbox{$\scriptstyle 1$}
\kern-0.1em/\kern-0.15em \lower0.25ex\hbox{$\scriptstyle
2$}},{\raise0.5ex\hbox{$\scriptstyle 3$} \kern-0.1em/\kern-0.15em
\lower0.25ex\hbox{$\scriptstyle 2$}}; -
{\raise0.5ex\hbox{$\scriptstyle {r^2 }$} \kern-0.1em/\kern-0.15em
\lower0.25ex\hbox{$\scriptstyle {4\theta }$}}} \right), \nonumber\\
\label{NCunp80}
\end{eqnarray}
where $\Phi \left( {d_U  - {\raise0.5ex\hbox{$\scriptstyle 1$}
\kern-0.1em/\kern-0.15em \lower0.25ex\hbox{$\scriptstyle
2$}},{\raise0.5ex\hbox{$\scriptstyle 3$} \kern-0.1em/\kern-0.15em
\lower0.25ex\hbox{$\scriptstyle 2$}}; -
{\raise0.5ex\hbox{$\scriptstyle {r^2 }$} \kern-0.1em/\kern-0.15em
\lower0.25ex\hbox{$\scriptstyle {4\theta }$}}} \right)$ is the
confluent hypergeometric function. From this expression, it should
be clear that the interaction energy is regular at the origin (see
Fig.$1$). Note that in Fig.$1$ we have defined $V= \frac{{q^2
}}{{\left( {2\pi } \right)^{2d_U  - 1} }}\frac{1}{{\Lambda ^{2d_U  -
2} }} \frac{1}{{\left( \theta  \right)^{d_U  -
{\raise0.5ex\hbox{$\scriptstyle 1$} \kern-0.1em/\kern-0.15em
\lower0.25ex\hbox{$\scriptstyle 2$}}} }}\frac{{\Gamma \left( {d_U  +
{\raise0.5ex\hbox{$\scriptstyle 1$} \kern-0.1em/\kern-0.15em
\lower0.25ex\hbox{$\scriptstyle 2$}}} \right)}}{{\Gamma \left( {2d_U
} \right)}} V[x]$. Notice that $\Phi \left( {a,c;z} \right) \to 1 +
\frac{a}{c}z$ as $z \to 0$. Thus, we obtain a constant potential at
the origin. In other words, the presence of $\theta$ gets rid of
short distances divergences. On the other hand, as long as $\Phi
\left( {a,c;z} \right) \to \frac{{\Gamma \left( c \right)}}{{\Gamma
\left( {c - a} \right)}}\left( { - z} \right)^{ - a}  +
\frac{{\Gamma \left( c \right)}}{{\Gamma \left( a \right)}}e^z z^{a
- c}$ as $z \to \infty$, we recover the long range un-particle
potential  $V \sim \frac{1}{{L^{2d_U  - 1} }}$, while that
short-distance effects are exponentially suppressed ${\mathcal
O}\left( {e^{ - \frac{{L^2 }}{{4\theta }}} } \right)$.

\begin{figure}[h]
\begin{center}
\includegraphics[scale=0.98]{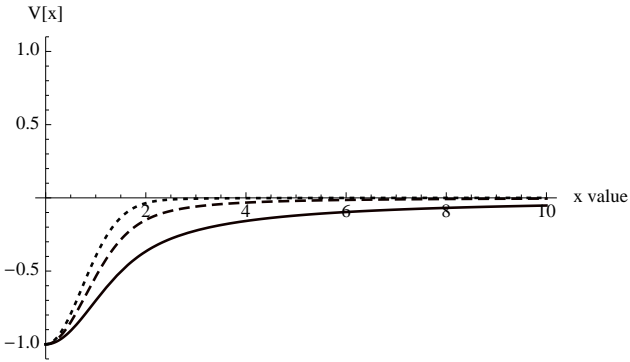}
\end{center}
\caption{\small The potential $V$, as a function of $x
= {\textstyle{r \over {2\sqrt \theta  }}}$. $d_U  = 1.1, 1.5$ and
$1.9$, are represented by the solid, dashed and dotted lines.
\label{fig1}}
\end{figure}

\section{Final Remarks}

To conclude, let us our work in its proper perspective.
As was showed, our analysis revealed that even if the Higgs
field takes a non-vanishing v.e.v., gauge interaction keeps its long
range character leading to an effective gauge symmetry restoration.
Besides, we have considered the effect of a minimal length on a
physical observable in the un-particle sector. Interestingly enough,
expression (\ref{NCunp80}) displays a marked departure from its
commutative counterpart. Nevertheless, the above profile potential
is similar to that encountered in ordinary space-time as $L
\to\infty$. Therefore, we have provided a new connection between
effective models. 

Finally, we would like to comment  our results from the vantage point
of the space-time "fractalization" outlined in \cite{Nicolini:2010bj}, 
where the concept of un-spectral dimension was introduced for the
first time. The authors in \cite{Nicolini:2010bj} proposed a new scenario
merging the conjectured existence of both  un-particles and minimal length
in the space-time fabric. In this case, the minimal length plays a role
which is similar to the critical temperature marking different phases
of some condensed matter system. The use of un-particle probes
allows the access to a trans-planckian regime which is forbidden to
ordinary matter. In this ultimate  phase the un-spectral space-time 
dimension is $ D_u = 2d_u-2$. We notice that the distance dependence
of the interaction potential (31) is just $1/L^{D_u}$, i.e. test charges
are sensitive to the un-spectral dimension. We consider this behavior
as a self-consistency check with the model discussed in
\cite{Nicolini:2010bj}.

\acknowledgments
This work was partially supported by Fondecyt (Chile) Grant 1130426 (PG).


\begin{thebibliography}{0}
\bibitem{Georgi:2007ek} H.~Georgi, Phys.\ Rev.\ Lett.\  {\bf 98}, (2007) 221601.

\bibitem{Georgi:2007si} H.~Georgi, Phys.\ Lett.\ B {\bf 650}, (2007) 275.

\bibitem{Nakayama:2007qu} Y.~Nakayama, Phys.\ Rev.\ D {\bf 76}, (2007) 105009.

\bibitem{Zhang:2007ih} H.~Zhang, C.~S.~Li and Z.~Li, Phys.\ Rev.\ D {\bf 76}, (2007) 116003.

\bibitem{Liao:2007ic} Y.~Liao and J.~-Y.~Liu, Phys.\ Rev.\ Lett.\  {\bf 99}, (2007) 191804.

\bibitem{Kikuchi:2007az} T.~Kikuchi and N.~Okada, Phys.\ Lett.\ B {\bf 665}, (2008) 186.
 
\bibitem{Deshpande:2007jy} N.~G.~Deshpande, X.~-G.~He and J.~Jiang, Phys.\ Lett.\ B {\bf 656}, (2007) 91.

\bibitem{Mureika:2007nc} J.~R.~Mureika, Phys.\ Lett.\ B {\bf 660}, (2008) 561.

\bibitem{Mureika:2008dx} J.~R.~Mureika, Phys.\ Rev.\ D {\bf 79} (2009) 056003.

\bibitem{Mureika:2009ii} J.~R.~Mureika, Int.\ J.\ Theor.\ Phys.\  {\bf 51} (2012) 1259.

\bibitem{Mureika:2010je} J.~R.~Mureika and E.~Spallucci, Phys.\ Lett.\ B {\bf 693}, (2010) 129. 

\bibitem{Gaete:2010sp} P.~Gaete, J.~A.~Helayel-Neto and E.~Spallucci,
Phys.\ Lett.\ B {\bf 693}, (2010) 155.
 
\bibitem{Ho:2009zv} C.~M.~Ho and Y.~Nakayama, JHEP {\bf 0905}, (2009) 081.

\bibitem{Mohanta:2007uu} R.~Mohanta and A.~K.~Giri, Phys.\ Rev.\ D {\bf 76}, (2007)  057701. 

\bibitem{Alan:2007ss} A.~T.~Alan and N.~K.~Pak, Europhys.\ Lett.\  {\bf 84}, (2008) 11001. 

\bibitem{Aliev:2009sd} T.~M.~Aliev, M.~Frank and I.~Turan, Phys.\ Rev.\ D {\bf 80}, (2009) 114019.

\bibitem{Kumar:2007af} M.~C.~Kumar, P.~Mathews, V.~Ravindran and A.~Tripathi, Phys.\ Rev.\ D {\bf 77}, (2008) 055013.

\bibitem{Frassino}  
A.~M.~Frassino, P.~Nicolini and O.~Panella, ``Un-Casimir effect'',
arXiv:1311.7173 [hep-ph].

\bibitem{Fox:2007sy} P.~J.~Fox, A.~Rajaraman and Y.~Shirman, Phys.\ Rev.\ D {\bf 76}, (2007) 075004.

\bibitem{Delgado:2007dx} A.~Delgado, J.~R.~Espinosa and M.~Quiros,
 JHEP {\bf 0710}, (2007) 094.
  
\bibitem{Kikuchi:2007qd} T.~Kikuchi and N.~Okada, Phys.\ Lett.\ B {\bf 661}, (2008) 360.


\bibitem{Sannino:2009za} F.~Sannino, Acta Phys.\ Polon.\ B {\bf 40}, (2009) 3533.

\bibitem{AmelinoCamelia:2002wr} G.~Amelino-Camelia, Nature {\bf 418},
(2002) 34.

\bibitem{Jacobson:2002hd}  T.~Jacobson, S.~Liberati and D.~Mattingly,
Phys.\ Rev.\ D {\bf 67}, (2003) 124011.

\bibitem{Konopka:2002tt} T.~J.~Konopka and S.~A.~Major,
 New J.\ Phys.\  {\bf 4}, (2002) 57.
  
\bibitem{Hossenfelder:2006cw}  S.~Hossenfelder, Phys.\ Rev.\ D {\bf 73}, (2006) 105013.

\bibitem{Nicolini:2008aj} P.~Nicolini, Int.\ J.\ Mod.\ Phys.\ A {\bf 24}, (2009) 1229.

\bibitem{Euro} P.~Nicolini, A.~Smailagic and E.~Spallucci, 
Phys.\ Lett.\ B {\bf 632}, (2006) 547. 

\bibitem{Witten} M. B. Green, J. H. Schwarz and E. Witten: Superstring Theory.
Cambridge University Press, Cambridge (1987).

\bibitem{Magueijo:2001cr}  J.~Magueijo and L.~Smolin,
Phys.\ Rev.\ Lett.\  {\bf 88}, (2002) 190403.

\bibitem{Hossenfelder:2003jz}  S.~Hossenfelder, M.~Bleicher, S.~Hofmann,
J.~Ruppert, S.~Scherer and H.~Stoecker, Phys.\ Lett.\ B {\bf 575}, (2003) 85.

\bibitem{Nicolini:2010bj} P.~Nicolini and E.~Spallucci, Phys.\ Lett.\ B {\bf 695}, (2011) 290.

\bibitem{Sprenger:2012uc}
M.~Sprenger, P.~Nicolini and M.~Bleicher, Eur.\ J.\ Phys.\  {\bf 33}, (2012) 853.

\bibitem{Witten:1985cc} E.~Witten, Nucl.\ Phys.\  B {\bf 268}, (1986) 253.

\bibitem{Seiberg:1999vs} N.~Seiberg and E.~Witten, JHEP {\bf 9909}, (1999) 032. 

\bibitem{Douglas:2001ba} M.~R.~Douglas, N.~A.~Nekrasov, Rev.\ Mod.\ Phys.\  {\bf 73},
(2001) 977-1029.

\bibitem{Szabo:2001kg} R.~J.~Szabo, Phys.\ Rept.\  {\bf 378}, (2003) 207-299.

\bibitem{Gomis:2000sp} J.~Gomis, K.~Kamimura and T.~Mateos, JHEP {\bf 0103},
(2001) 010.

\bibitem{Bichl:2001nf} A.~A.~Bichl, J.~M.~Grimstrup, L.~Popp, M.~Schweda
and R.~Wulkenhaar, Int.\ J.\ Mod.\ Phys.\  A {\bf 17}, (2002) 2219.

\bibitem{Smailagic:2003rp} A.~Smailagic and E.~Spallucci, J.\ Phys.\ A  {\bf 36}, (2003) L517. 

\bibitem{Smailagic:2003yb} A.~Smailagic and E.~Spallucci, J.\ Phys.\ A  {\bf 36}, (2003) L467. 

\bibitem{Smailagic:2004yy} A.~Smailagic and E.~Spallucci, J.\ Phys.\ A  {\bf 37}, (2004) 1 [Erratum-ibid.\  A {\bf 37}, (2004) 7169].

\bibitem{Hammou:2001cc} A.~B.~Hammou, M.~Lagraa and M.~M.~Sheikh-Jabbari, Phys.\ Rev.\ D {\bf 66}, (2002) 025025.

\bibitem{Gaete:2012cs}{}P.~Gaete and E.~Spallucci, {\bf ``Gauge symmetry restoration in the un-particle vacuum''}, {}arXiv:1205.2248 [hep-th].
  
\bibitem{Gaete:2008wg} P.~Gaete and E.~Spallucci, Phys.\ Lett.\ B {\bf 661}, (2008) 319.

\bibitem{Krasnikov:2007fs} N.~V.~Krasnikov, Int.\ J.\ Mod.\ Phys.\ A {\bf 22}, (2007) 5117.

\bibitem{Gaete:2008aj} P.~Gaete and E.~Spallucci, Phys.\ Lett.\ B {\bf 668}, (2008) 336.

\bibitem{Gaete:2011ka} P.~Gaete and E.~Spallucci, J.\ Phys.\ A {\bf 45}, (2012) 065401.

\bibitem{Gaete:2012yu} P.~Gaete, J.~ Helayel-Neto and E.~Spallucci,
J.\ Phys.\ A {\bf 45}, (2012) 215401.

\bibitem{Smailagic} E.~Spallucci, A.~Smailagic and P.~Nicolini,
 Phys.\ Rev.\ D {\bf 73}, (2006) 084004. 

\bibitem{Moffat} L.~Modesto, J.~W.~Moffat and P.~Nicolini,
Phys.\ Lett.\ B {\bf 695}, (2011) 397. 

\bibitem{Nicolini} P.~Nicolini,
``Nonlocal and generalized uncertainty principle black holes'',
arXiv:1202.2102 [hep-th].
\end{thebibliography}
\end{document}